\documentclass[%
preprint,
 amsmath,amssymb,
 aps,
]{revtex4-1}

\usepackage{graphicx}
\usepackage{dcolumn}
\usepackage{amsmath}
\usepackage[bbgreekl]{mathbbol}
\usepackage{amssymb}
\usepackage{bm}
\usepackage{subfigure}
\usepackage{slashed}
\usepackage{hyperref}
\usepackage{mathrsfs}

\usepackage{tikz}
\usetikzlibrary{shapes,shadows,arrows,positioning,calc,intersections}
\tikzstyle{block}=[draw,rectangle, text centered, node distance=7em]
\tikzstyle{line}=[draw,-latex]

\usepackage{bm}

\def\p{\mbox{\boldmath$\displaystyle\mathbf{p}$}}

\def\0{\mbox{\boldmath$\displaystyle\mathbf{0}$}}

\def\x{\mbox{\boldmath$\displaystyle\mathbf{x}$}}

\begin{document}


\title{A QFT-induced phase  in neutrino flavour oscillations}

\author{Dharam~Vir~Ahluwalia$\empty^{a,b,}$}
\email{d.v.ahluwalia@iitg.ac.in}
\affiliation{$\empty^{a}$ Department of Physics,
Indian Institute of Technology Guwahati,\\
Guwahati - 781 039, Assam, India}

\affiliation{$\empty^{b}$ Centre for the Studies of the Glass Bead Game\\
Chaugon, Bir, Himachal Pradesh 176 077, India}

\author{Cheng-Yang Lee}
\email{cy.lee@manipal.edu}
\affiliation{
Manipal Centre for Natural Sciences, Manipal University,\\
Manipal, Karnataka 576104, India
}

\date{24 May 2017}

\begin{abstract}
In the extended Standard Model of particle physics, 
each neutrino mass eigenstate is predicted to have a tiny but non vanishing magnetic moment induced by quantum field theoretic  corrections. 
These QFT-induced magnetic momenta depend linearly on masses of the underlying mass eigenstates  with a proportionality  constant 
${3eG_{F}}/{(8\sqrt{2}\pi^{2})} $. As a consequence when neutrinos are embedded in an environment containing magnetic fields the flavour oscillations get a contribution from the induced relative phases.


\end{abstract}

\maketitle



If Pontecorvo formalism to understand neutrino oscillations is correct, then neutrinos provide concrete evidence for physics beyond the Standard Model (SM). The observations of flavour oscillations then mean that neutrinos produced through electroweak processes must be in linear superposition of  at least two non-vanishing masses contrary to the prediction of the SM -- the lowest mass eigenstate may be massless, or carry a non-zero mass.

Due to quantum field theoretic corrections (QFT), various extensions of the SM predict that neutrinos should have small but non-zero magnetic moments ~\cite{Marciano:1977wx,Lee:1977tib,Pal:1981rm,Schechter:1981hw,Shrock:1982sc,Masood:1992dg,
Masood:1995aq,Dvornikov:2004sj,Bell:2005kz,Bell:2007nu}. For the Dirac neutrinos, the magnetic moments corresponding to the individual mass eigenstates are given by
\begin{eqnarray}
\bbmu_{i}&=&\frac{3eG_{F}m_{i}}{8\sqrt{2}\pi^{2}} \nonumber\\
&=&1.85\times 10^{-27}\left(\frac{m_{i}}{\mbox{eV}}\right)\left(\frac{\mbox{eV}}{\mbox{Gauss}}\right).
\end{eqnarray}

The physics of the neutrino magnetic moment have been extensively studied in the forms of spin precession in magnetic fields and the helicity-flipping scattering amplitudes~\cite{Cisneros:1970nq,Fujikawa:1980yx,Dvornikov:2002rs,Studenikin:2004tv,Studenikin:2004bu,Giunti:2008ve,
Giunti:2010zz,Broggini:2012df,Giunti:2014ixa,
Dobrynina:2016rwy,Fabbricatore:2016nec,Studenikin:2016ykv}. In this letter, we report a new type of $\bbmu$-induced flavour oscillation that to the best of our knowledge, has not been considered. The crucial observation lies in the fact that since neutrinos are predicted to have non-zero magnetic moments, the energies, and hence the induced phases, of the mass eigenstates are shifted when travelling through magnetic fields. Moreover, this shift varies with the mass of the mass eigenstate in the superposition that defines the neutrino flavour. This circumstance adds a new element to amplitudes, and hence probabilities, for neutrino flavour oscillations. 

A simple analysis  shows that the kinematically induced phase
\begin{equation}
\varphi_{ji} \equiv\frac{\Delta m_{ji}^{2}L}{4E}=1.27\left(\frac{\Delta m_{ji}^{2}}{\mbox{eV}^{2}}\right)
\left(\frac{L}{\mbox{km}}\right)\left(\frac{\mbox{GeV}}{E}\right),
\end{equation}
gets an additive QFT-induced correction
\begin{eqnarray}
 \varphi_{ji}^{\mbox{\tiny{new}}} && \equiv \varphi_{ji}\left(\frac{3eG_{F}B}{4\sqrt{2}\pi^{2}}\right)\cos(\eta) \nonumber \\
&&= \varphi_{ji} \times 3.7\times 10^{-27} \cos(\eta)\,
\left(\frac{B}{\mbox{Gauss}}\right).
\end{eqnarray}
Where $\eta$ is the angle between the magnetic field (assumed constant and uniform) and the direction of propagation, and $\Delta m^2_{ji} = m^2_j-m^2_i$ is the usual mass-squared difference between the $i$th and $j$th mass eigenstates.
\vspace{11pt}

\noindent
\textit{Some details of the analysis} \textemdash~
The just quoted results require a calculation of the interaction energy of the magnetic moment of a Dirac  mass eigenstate with an external magnetic field. We do this calculation for a uniform and constant magnetic field. The effective coupling of the neutrino to the classical electromagnetic field due to the magnetic moment is
\begin{equation}
V=\int d^{3}x\, J^{\mu}_{\mbox{\tiny{eff}}} A_{\mu}
\end{equation}
where $J^{\mu}_{\mbox{\tiny{eff}}}(x)$ is the current. We take its matrix elements to be of the form~\cite{Giunti:2014ixa}
\begin{widetext}
\begin{eqnarray}
\langle p',\sigma'|J^{\mu}_{\mbox{\tiny{eff}}}(x)|p,\sigma\rangle&=&e^{-i(p-p')\cdot x}\langle p',\sigma'|J^{\mu}_{\mbox{\tiny{eff}}}(0)|p,\sigma\rangle\nonumber\\
&\equiv &\frac{1}{2(2\pi)^{3}\sqrt{EE'}}e^{-i(p-p')\cdot x}\bar{u}(\p',\sigma')\Gamma^{\mu}(p',p)u(\p,\sigma)
\end{eqnarray}
where $\Gamma^{\mu}(p',p)\equiv -i\bbmu \sigma^{\mu\nu}(p-p')_{\nu}$ with $\bbmu$ being the neutrino magnetic moment and $\sigma^{\mu\nu}=(i/2)[\gamma^{\mu},\gamma^{\nu}]$, and the remaining symbols have their usual meaning. The normalization factor comes from our choice of expansion for the fermionic field
\begin{equation}
\psi(x)=(2\pi)^{-3/2}\int\frac{d^{3}p}{\sqrt{2E}}\sum_{\sigma}[e^{-ip\cdot x}u(\p,\sigma)a(\p,\sigma)
+e^{ip\cdot x}v(\p,\sigma)b^{\dag}(\p,\sigma)]. 	
\end{equation}
\end{widetext}
Since we are interested in neutrinos propagating through a constant magnetic field, the potential is time-independent $A^{\mu}=A^{\mu}(\x)$. We work in the  temporal gauge and set $A^{0}(x)=0$.  With these facts noted, we obtain
\begin{equation}
\langle p',\sigma'|V|p,\sigma\rangle=-\frac{\bbmu}{E}\bar{u}(\p',\sigma')\,
J^{ij}u(\p,\sigma)F_{ji}\delta^{3}(\p'-\p)\label{eq:v}
\end{equation}
where $J^{ij}=(i/4)[\gamma^{i},\gamma^{j}]$ and $F_{ij}=\partial_{i}A_{j}-\partial_{j}A_{i}$. Using the identity $F_{ij}=-\epsilon_{ijk}B^{k}$, equation~(\ref{eq:v}) simplifies to
\begin{equation}
\langle p',\sigma'|V|p,\sigma\rangle=-\left(\frac{\bbmu}{E}\right)
\bar{u}(\p',\sigma')(\mathbf{J\cdot B})
u(\p,\sigma)\delta^{3}(\p'-\p).
\end{equation}

Generally, the neutrinos produced in the weak interactions are relativistic helicity eigenstates. In this case, we have
\begin{equation}
\bar{u}(\p,\sigma)(\mathbf{J\cdot B})u(\p,\sigma)\approx 2m\sigma\,\widehat{\p}\cdot\mathbf{B}.
\end{equation}
Therefore, by taking $p=p'$, $\sigma=\sigma'$ and normalising the matrix element, the shift in energy is given by
\begin{equation}
V_{\tiny{\mbox{rel}}}\equiv\frac{\langle p,\sigma|V_{\tiny{\mbox{rel}}}|p,\sigma\rangle}{\langle p,\sigma|p,\sigma\rangle}\approx\-2\sigma \left(\frac{m}{E}\right)(\bbmu\widehat{\p}\cdot\mathbf{B}).\label{eq:e1}
\end{equation}
For completeness, in the non-relativistic limit where $\p\sim\0$, we have
\begin{equation}
\bar{u}(\0,\sigma)(\mathbf{J\cdot B})u(\0,\sigma)=2m\sigma B_{z}
\end{equation}
so
\begin{equation}
V_{\tiny{\mbox{non-rel}}}\equiv\frac{\langle k,\sigma|V_{\tiny{\mbox{non-rel}}}|k,\sigma\rangle}{\langle k,\sigma|k,\sigma\rangle}\approx-2\sigma \bbmu B_{z}\label{eq:e2}
\end{equation}
where $k^{\mu}=(m,\0)$ and $B_{z}$ is the magnetic field along the $z$-axis. The spin of the state $|k,\sigma\rangle$ is aligned with the $z$-axis.

\noindent\textit{Concluding remarks} \textemdash~
A class of neutron stars called magnetar can have magnetic fields of the order of $10^{16}$ Gauss~\cite{Tiengo:2013gsa}. This falls some eleven orders of magnitude too short to effect neutrino oscillations in the present cosmic epoch.  However, with cosmic expansion the magnetic field scales as
$
B_e = B_p \left({a_p}/{a_e}\right)^2
$
where $a$ is the scale factor, while  $e$ refers to the end of inflation and $p$ denotes the present epoch. Trivedi and  Subramanian argue that CMB observations provide an upper limit for $B_p$ to be 
$\alt 0.05$ nano-Gauss~\cite{PhysRevD.89.043523}. It is not clear to us as to how far back we may extrapolate the naive increase in the magnetic field to early universe to entertain a dramatic change in neutrino oscillations.
\vspace{11pt}

\noindent\textit{Acknowledgement}~\textemdash
C.~Y.~L would like to thank the generous hospitality offered by the Department of Physics and Astronomy of the University of Canterbury where part of this work was completed.  

\providecommand{\href}[2]{#2}\begingroup\raggedright\endgroup


\end{document}